\documentclass[10pt,conference]{IEEEtran}

\ifCLASSINFOpdf
  \usepackage[pdftex]{graphicx}
  \graphicspath{{./images/}}
\else
  \usepackage[dvips]{graphicx}
  \graphicspath{{./images/}}
\fi

\usepackage[cmex10]{amsmath}
\usepackage[linesnumbered,ruled,vlined,longend]{algorithm2e}
\usepackage{siunitx}
\usepackage{array}

\usepackage[tight,footnotesize]{subfigure}
\usepackage{caption}
\usepackage{subcaption}

\usepackage{url}

\usepackage[usenames,dvipsnames,svgnames,table]{xcolor}
\usepackage{amsfonts}
\usepackage{amsthm}
\usepackage{mathrsfs}
\usepackage{adjustbox}
\usepackage{multirow}

\theoremstyle{definition}

\begin{document}

\title{{A streaming algorithm and hardware accelerator for top-$K$ flow detection in network traffic}}
\author{
		\IEEEauthorblockN{Carolina Gallardo-Pavesi\IEEEauthorrefmark{1}, Yaime Fernández\IEEEauthorrefmark{1}, Javier E. Soto\IEEEauthorrefmark{1}, Cecilia Hernández\IEEEauthorrefmark{2}\IEEEauthorrefmark{3}, and Miguel Figueroa\IEEEauthorrefmark{1}\IEEEauthorrefmark{4}
       }
		\IEEEauthorblockA{
                 \IEEEauthorrefmark{1}Department of Electrical Engineering, \IEEEauthorrefmark{2}Department of Computer Science -  Universidad de Concepci\'on, Concepci\'on, Chile\\  \IEEEauthorrefmark{3}Center of Biotechnology and Bioengineering (CeBiB)\\ \IEEEauthorrefmark{4}Advanced Center for Electrical and Electronics Engineering (AC3E)\\
                e-mail: \{cgallardo2018, yfernandezj, javsoto, cecihernandez, miguel.figueroa\}@udec.cl
}
}


	\maketitle
	\begin{abstract}
	    Identifying the largest $K$ flows in network traffic is an important task for applications such as flow scheduling and anomaly detection, which aim to improve network efficiency and security. However, accurately estimating flow frequencies is challenging due to the large number of flows and increasing network speeds. Hardware accelerators are often used in this endeavor due to their high computational power, but their limited amount of on-chip memory constrains their performance. Various sketch-based algorithms have been proposed to estimate properties of traffic such as frequency, with lower memory usage and theoretical bounds, but they often under perform with the skewed distribution of network traffic. In this work, we propose an algorithm for top-$K$ identification using a modified TowerSketch and a priority queue array. Tested on real traffic traces, we identify the top-$K$ flows, with $K$ up to 32,768, with a precision of more than 0.94, and estimate their frequency with an average relative error under 1.96\%. We designed and implemented an accelerator for this algorithm on an AMD Virtex U280 UltraScale+ FPGA, which processes one packet per cycle at 392~MHz, reaching a minimum line rate of more than 200~Gbps.
	\end{abstract}
		
\section{Introduction}




In network traffic measurement, a flow is a sequence of packets with common attributes such as source and destination IP addresses, source and destination ports, and the communication protocol. Notably, the top-$K$ flows are the $K$ flows with the largest number of packets (frequency) or size (payload) \cite{zhang2024jigsaw}. Real-time identification of top-$K$ flows is critical for network performance and security, supporting applications such as flow scheduling~\cite{yu2019dshark}, load balancing~\cite{sivaraman2016programmable}, anomaly detection~\cite{li2006detection}, and heavy-hitter and heavy-change detection~\cite{harrison2018network, schweller2004reversible}.



Identifying the top-$K$ flows requires simultaneously tracking the identifiers (the set of common attributes) and size of each flow~\cite{zhang2024jigsaw}, using a counter per flow to accumulate its frequency or payload.
Moreover, modern network monitoring requires processing packets at increasingly high speeds to keep up with faster data links, which is challenging to do using general-purpose processors. To address this problem, dedicated hardware accelerators and specialized programmable switches have been developed to process large data streams with high throughput and low latency. Programmable switches use fast application-specific integrated circuits (ASICs) to provide very high throughput, but their programming model limits the range of algorithms that can be efficiently implemented on them~\cite{ding2020estimating}. Dedicated hardware accelerators based on field-programmable gate arrays (FPGAs) are more flexible and can achieve high performance~\cite{tong2017sketch, lai2019sketch}, although they normally achieve lower performance compared to an equivalent ASIC. 

Programmable switches and hardware accelerators have limited on-chip memory resources; therefore, their performance is often limited by their bandwidth to external memory. This makes it impossible to maintain a counter for each possible flow and update them at line rate without significant packet loss. An efficient way to solve this problem is to use probabilistic data structures such as sketches, which allow us to estimate properties of large amounts of data with high accuracy, low memory usage, and theoretical error bounds. These algorithms are especially attractive for hardware accelerators because they can process the data in a single pass and exhibit a high degree of fine-grained parallelism that can be exploited in hardware. 
Classic sketches for frequency and size estimation include CountMin~\cite{cormode2011approximating}, CountMin with Conservative Updates (CountMin-CU)~\cite{goyal2012sketch}, and CountSketch~\cite{ charikar2002finding}, with more recent proposals such as Elastic~\cite{yang2018elastic}  and TowerSketch~\cite{yang2023sketchint} addressing the skewed distribution of network traffic. 
Although these sketches have often been used in hardware accelerators~\cite{tong2017sketch, soto2021high, fernandez2023streaming, yang2018elastic, yang2023sketchint}.
they require a large amount of memory to accurately estimate the frequencies, often resulting in significant overestimation of smaller counts. This can negatively affect the accuracy of the detection of the top-$K$ flows, especially because the distribution of network traffic is highly skewed, so there is a large difference between the largest and the smaller flows in the top-$K$~\cite{barabasi2013network}.

In this paper, we present an algorithm to estimate the top-$K$ flows and their frequencies, using a modified version of the TowerSketch and an approximate priority queue that can be updated in constant time. We also present the architecture and implementation of a hardware accelerator for this algorithm. Using CAIDA traces~\cite{CAIDA}, 
the accelerator detects the top-$K$ flows with a precision of more than 0.94 for values of $K$ up to 32,768 flows. It also estimates their frequencies with an average relative error of less than 1.96\%. Implemented on an AMD Virtex XCU280 UltraScale+ FPGA, the accelerator runs at 392~MHz and processes one packet per cycle, allowing it to operate at line rates of at least 200 gigabits per second (Gbps).


\section{Related work}
 
\subsection{Algorithms for top-$K$ estimation}

Several streaming algorithms have been proposed for top-$K$ detection in network traffic. Some use a count sketch~\cite{yang2024cts, zheng2024detecting} to estimate the flow size and a separate data structure to store the top frequencies, some use fixed-size counters~\cite{yang2019heavykeeper} to keep track of the frequencies, and others combine both~\cite{shi2023cuckoo}.

In~\cite{yang2024cts}, authors proposed the CTS sketch to identify top-$K$ flows. It uses a data structure on edge switches to record large flows and a global flow table with a voting mechanism on the centralized controller~\cite{yang2018elastic} to select the top-$K$ flows. Using the complete 5-tuple as the flow identifier, they achieve 89\% precision when detecting top-$100$ flows in synthetic traces. 
HeavyKeeper~\cite{yang2019heavykeeper} uses a hash table and an exponential decay, achieving a precision of more than 94\% when detecting the top-$1,000$ flows in the CAIDA~\cite{CAIDA} 2016 dataset with a memory size of 100~KB. Flows are identified only by their source and destination IP addresses. Other works, such as Cuckoo Counter (CC)~\cite{shi2023cuckoo} combine counter-based structures and sketches. The authors used two arrays, each containing buckets with multiple entries of different sizes. They use an additional heap that records the frequency of a flow when it is first inserted. 
They achieve 95\% precision when detecting the top-$1,000$ flows in CAIDA, IMC and Real-Life, also using only the source and destination IP addresses. 
Recently, Cao \emph{et al.}~\cite{cao2024bubble} proposed BubbleSketch, which is composed of two arrays with entries of varying sizes, differentiating between large and small flows.  They show 100\% precision in CAIDA 2016 traces when $500 \leq K \leq 5,000$, but the arrays must be sorted on line as new elements are inserted.

The precision of top-$K$ estimation depends strongly on the count sketch used to estimate the flow frequencies. In skewed distributions, which are typical in network traffic, only a small minority of flows have high frequencies. TowerSketch~\cite{yang2023sketchint} has recently been proposed to exploit this property by assigning a larger number of small counters to flows with smaller frequencies and a smaller number of wider counters to the largest flows. The structure of the sketch is similar to CountMin, with rows of the same size but different number of counters. Using CAIDA traces and the source IP address as flow identifier, they achieve an absolute average error of 0.3 using the CountMin update and estimation criterion and 0.02 using CountMin-CU.






\subsection{Hardware accelerators using sketches}

Finding the top-$K$ elements is a particular case of finding heavy hitters, which requires identifying the items in the processed stream whose value exceeds a user-supplied threshold~\cite{ngo2024cuckoo}.  
In~\cite{zazo2017single}, the authors proposed a hardware accelerator to identify the source IP addresses that generate the most packets. They combine a count sketch, a priority queue, and a network packet parser and implement them on an AMD VCU108 FPGA. The on-chip queue is a sorted list of the top‐$15$ items. The maximum clock frequency achieved by their implementation is 322~MHz, allowing them to process more than 26.8 million packets per second at a line rate of 100~Gbps. The implementation was validated using CAIDA traces~\cite{CAIDA}, achieving a count estimate average relative error (ARE) of 1.29\%, but only in the top-$15$ flows. 

In~\cite{saavedra2018heavy}, we proposed an accelerator to detect heavy hitters in genomic and network traffic data. The accelerator estimates the frequency of each element with a CountMin-CU sketch and selects the heavy hitters using a user-supplied threshold. We use only the source IP address as a flow identifier. It runs on an AMD XC7K325T FPGA and operates with a 300~MHz clock. The $4\times2^{14}$-bucket sketch is implemented with on-chip memory. On the same line, Soto \emph{et al.}~\cite{soto2020hardware} proposed a hardware accelerator to estimate the entropy of only the top-$K$ flows. It combines a priority queue array (PQA) of 10,240 elements and a CountMin-CU sketch to estimate the flow entropy, which are identified only by their source IP address. Implemented on an AMD ZCU102 FPGA, the accelerator operates at a frequency of 354~MHz, and supports line rates of at least 181 Gbps. 
In a later work~\cite{soto2021high, fernandez2023streaming}, we used the top-$8,192$ flows to estimate the statistical traffic distribution, substantially improving the quality of the entropy estimates and using the complete 5-tuple to identify the flows. 


The Jigsaw Sketch was proposed in~\cite{zhang2024jigsaw} to detect top-$K$ flows. The sketch has two stages: First, a bucket array with a probabilistic replacement strategy is used to identify large flows. The second stage stores the flow identifiers of the candidate top-$K$ flows in an auxiliary list. They implemented the algorithm on a NetFPGA-1G-CML using 200 KB of memory and operating at 115.55 MHz. With CAIDA traces from 2016 and 2019, as $K$ increases from 1,000 to 5,000, the average relative error increases from 0.3\% to 1.4\% and from 0.1\% to 0.8\%, respectively, with a precision greater than 0.96.



In~\cite{yang2023sketchint}, the authors implemented TowerSketch on an SmartNIC with an AMD VC709 FPGA.  The sketch uses 3 rows, each containing counters of 8, 16 and 32 bits, respectively. It uses the CountMin update and estimation algorithm, that is, with each update, it increments all the selected counters and uses the minimum value as the frequency estimate. The total sketch size is 192~KB and runs at 365~MHz, using a pipeline that processes a new packet per clock cycle. 


      



     


 \section{Algorithms}

We aim to estimate the number of packets (frequency) in the top-$K$ flows, using the complete 5-tuple as a flow identifier: Source and destination address, source and destination port, and protocol. Algorithm~\ref{alg:general} depicts the general algorithm for top-$K$ detection. During an observation interval, we insert each incoming packet and its flow identifier into a counting sketch to estimate the flow frequency. 
Our algorithm uses a variation of the TowerSketch~\cite{yang2023sketchint} to count the number of packets in each flow. When a packet is inserted, we update the sketch and produce a frequency estimate, which is then inserted into an approximate priority queue (PQA) of fixed size that tracks the largest flows. After the observation interval, we sort the contents of the PQA and retrieve the top-$K$ flows. Below, we describe each part of the algorithm in detail. 

\begin{algorithm} [tb]
        \SetKwInOut{Input}{input}
        \Input{$m$ and $d$, dimensions of the TowerSketch, 
        $L$ number of elements in the PQA}
        \SetKwFunction{Finsert}{.Insert}
        \SetKwFunction{Fpqainsert}{.Insert}
        \SetKwFunction{Fpqasort}{.Sort}
        \SetArgSty{} 
        T $\leftarrow$ new\texttt{ TowerSketch($m$,$d$)}
        
        P $\leftarrow$ new \texttt{PQA($L$)}\\
        \ForEach{packet}
            {
                f $\leftarrow$ packet.flow\_id
                
                est $\leftarrow$ T\Finsert{f}\tcp*[f]{Algorithm 2}
                \\ 
                P\Fpqainsert{f, est} \tcp*[f]{Algorithm 3}
                
            }
        P\Fpqasort{}
        
        \KwRet{$K$ most frequent flows from PQA} 
        \caption{\textbf{General Algorithm}}
        \label{alg:general}
 \end{algorithm}

 \subsection{TowerSketch with conservative updates}
  \label{sec:Tower}

TowerSketch~\cite{yang2023sketchint} is a probabilistic data structure for frequency estimation. Like traditional counting sketches such as CountMin and CountSketch, it is structured as $d$ rows of counters that are updated in each insertion. Unlike the others, TowerSketch is designed for skewed data distributions. It uses counters with different bit widths for each row so that each row uses the same amount of memory. Consequently, the sketch provides more small counters to record less frequent flows and fewer large counters to record the more frequent ones, which is congruent with the typical distribution of network traffic. Every row has $w_i$ counters of $\delta_i$ bits, with $i\in[1,d]$. The maximum value a counter can record is $2^{\delta_i} -2$. When a counter reaches $2^{\delta_i} -1$, it is considered overflowed and its value is interpreted as $+\infty$. For each new packet, its flow identifier is mapped onto one counter in each row using $d$ hash functions. With Conservative Updates~\cite{yang2023sketchint}, only the counters with the minimum value are incremented by one, and the new minimum value provides the current frequency estimate. In the original formulation of the sketch, the number of bits in each consecutive row is doubled with respect to the previous one.

       \begin{figure}[tb]
	\begin{center}
	\includegraphics[width=0.9\linewidth]{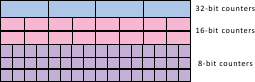}
		\caption{TowerSketch structure. The sketch uses six rows: three with 8-bit counters, two with 16-bit counters and one with 32-bit counters. Each row uses the same amount of memory.} 
                \label{towercu}
				\end{center}
\end{figure}  

As discussed in Section~\ref{sec:results}, we use 1-minute CAIDA traces to evaluate the performance of our algorithm and accelerator. We want to capture the set of top-$K$ flows responsible for at least half of the total traffic, which requires a minimum counter size of 8 bits. Therefore, a TowerSketch in its original formulation would use 3 rows with counters of 8, 16, and 32 bits. However, because of the highly skewed distribution of the traffic, it is better to increase even more the number of small counters. Therefore, we modify the sketch to use 6 rows in total: 3 rows of 8-bit counters, 2 rows of 16-bit counters, and 1 of 32-bit counters,  as shown in Fig.~\ref{towercu}. In Section \ref{sec:results}, we compare this approach to the original TowerSketch using the same amount of memory as our solution, that is, twice the number of counters in each row. 

Following Algorithm~\ref{alg:general}, for each incoming packet we update the sketch and produce an estimate of the corresponding flow frequency on line. Algorithm~\ref{insalg} shows the insertion and estimation algorithm for TowerSketch with conservative updates (Tower-CU), which has been shown to provide the best results~\cite{yang2023sketchint}. A 32-bit MurmurHash3 hash function with different seeds per row maps the flow identifier to one bucket (counter) for each row, using the number of bits needed to address each row according to its dimension $w_i$. Then, we find the minimum value stored in the counters. Since the overflowed buckets count as $+\infty$, we check that the buckets are not overflowed before updating the minimum value. Later, we add 1 to each bucket that has the minimum value and is not overflowed and write the buckets back into the sketch. The flow frequency estimate corresponds to the minimum value of the updated buckets that have not overflowed.

 \begin{algorithm} [tb]
     \SetKwInOut{Input}{input}\SetKwInOut{Output}{output}
     \SetKwProg{Fn}{\textbf{function}}{}{}
     \SetKwFunction{Fminval}{Findminval}
     \SetKwFunction{Fhash}{MurmurHash3}
     \SetKwFunction{Fest}{Insert}
     \SetArgSty{} 
     \Input{A packet belonging to flow with identifier $f$}
     \Output{\textit{est}, frequency estimation for $f$}
    
    $i\leftarrow 0$
     
     minval $\leftarrow 2^{32} -1 $
     
     \While{$i<d$}{
          \tcp{MurmurHash3 is a 32-bit hash function}
       hash $\leftarrow$  \Fhash{f, seeds[$i$]}
        
        idx[$i$] $\leftarrow$ hash $\& \ (w_i - 1)$
        
         \tcp{tower is the  TowerSketch storage}
       bucket[$i$] $\leftarrow$ tower[$i$][idx[$i$]]
        
        \tcp{$\delta_i$ is the bucket width in bits}
        \If{bucket[$i$] $<$ minval
        \textbf{and} \\ bucket[$i$] $\neq \ 2^{\delta_i}-1$}{
            minval $\leftarrow $bucket[$i$]
        }
        $i \leftarrow i+1$
    }

    $i\leftarrow 0$
    
    minval2 $\leftarrow 2^{32} -1 $
    
    \While{$i<d$}{
                
        \If{bucket[$i$] $=$ minval}{
            \If{bucket[$i$] $\neq \ 2^{\delta_i}-1$}{
                bucket[$i$] $\leftarrow$ bucket[$i$] $+1$;
                
                tower[$i$][idx[$i$]] $\leftarrow$ bucket[$i$]
            }
        }
        \If{bucket[$i$] $<$ minval2
        \textbf{and} \\ bucket[$i$] $\neq \ 2^{\delta_i}-1$}{
            minval2 $\leftarrow $bucket[$i$];
        }
        $i\leftarrow i+1$
    }
    
    \textit{est} $\leftarrow $ minval2
    
    \KwRet{\textit{est}}
    	
        \caption{\textbf{TowerSketch Insertion and Estimation}}
        \label{insalg}
 \end{algorithm}

        \subsection{Priority queue array}
        \label{sec:algorithm}

        To extract the top-$K$ flows, we need to insert each new flow frequency estimation provided by the sketch into a fixed-size priority queue that maintains the $K$ flows with the highest frequency. However, implementing a priority queue requires sorting the elements inside it every time a new insertion occurs, which requires a minimum of $\log_2{K}$ memory accesses per insertion. This expensive operation would limit accelerator throughput, especially for large values of $K$. 
        
        
        Instead, we use a priority queue array (PQA)~\cite{fernandez2023streaming} that approximates the behavior of a priority queue by using an array of $R$ priority queues of $S$ elements, where $K=R \times S$ and $S \ll R$. Each inserted element is mapped to one of the priority queues using a hash function. Since the hash function tries to distribute the flows uniformly among the queues, the data structure captures an important fraction of the elements of an actual priority queue. Moreover, because $S$ is a very small number, elements can be inserted into the queues in approximately constant time in a hardware implementation, using a combinational operation in a single clock cycle.
        
        Algorithm~\ref{pqains} shows the PQA insertion algorithm. For each new flow identifier, the PQA computes a 32-bit hash function, of which the $log_2 R$ least significant bits are used as an index to select one of the $S$-element priority queues. The remaining bits are used as a tag to distinguish the flow from the others that are mapped to the same queue. Thus, each flow in the queue is represented as a tuple formed by its estimated frequency and the tag. The incoming tag is compared to those of the flows stored in the queue, producing the flag \textit{found}, which indicates if the flow is already present in the queue, and its position $i$. If the flow was found, the PQA replaces its frequency with the incoming frequency \textit{est} if \textit{est} is greater than the stored frequency. Otherwise, if the flow is not present, it compares \textit{est} to the smallest frequency stored in the queue, keeping the flow with the highest frequency. Finally, the selected queue is sorted. Each of these operations can be performed in a single clock cycle in hardware when $S$ is small.
        

        \begin{algorithm}[tb] 
     \DontPrintSemicolon
	 	\SetKwInOut{Input}
         {input}\SetKwInOut{Output}{output}
         \SetKwFunction{Fhash}{MurmurHash3}
         \SetArgSty{} 
         \SetKwFunction{Ftag}{find\_tag}
		\Input{A flow identifier $f$ and \textit{est}, the frequency estimation for $f$}
        
             hash $\leftarrow$ \Fhash{f, seed}
             
            idx $\leftarrow$ hash $\& \ (R-1)$ 

            tag $\leftarrow$ hash $>> \log_2 (R)$
            
            pqa[idx].\Ftag{tag, i, found}
            
            \If{found}{
               \If{pqa[idx][i].count $<$ est}{
                pqa[idx][i].count $\leftarrow$ est
                }
            }
            \Else{
                \If{est $>$ pqa[idx][S-1].count}{
                    pqa[idx][S-1].count $\leftarrow$ est
                    
                    pqa[idx][S-1].tag $\leftarrow$ tag
                }            
            }
            \textit{sort}(pqa[idx])
            
	 	\caption{\textbf{PQA Insertion}}
		\label{pqains}

	 \end{algorithm}

The discussion above shows that the PQA can only store $S$ elements in each priority queue. However, collisions in the hash function can cause more than $S$ of the top-$K$ flows to be mapped to the same queue. As a result, the PQA loses these elements and degrades its precision when capturing the top-$K$ flows. To mitigate this problem, we increase the PQA size by adding additional elements to each priority queue (e.g., two additional elements when $S = 4$) to capture more flows. Once all the packets in the observation window are processed, we sort the PQA elements and extract the $K$ largest flows. Because this sorting is not performed as the packets are received and because $K$ is much smaller than the total number of flows, it can be performed in software in the control plane.

    \section{Hardware accelerator architecture}
    \label{sec:accelerator}
    
    \subsection{General architecture}
    \begin{figure}[tb]
	\begin{center}
	\includegraphics[width=1\linewidth]{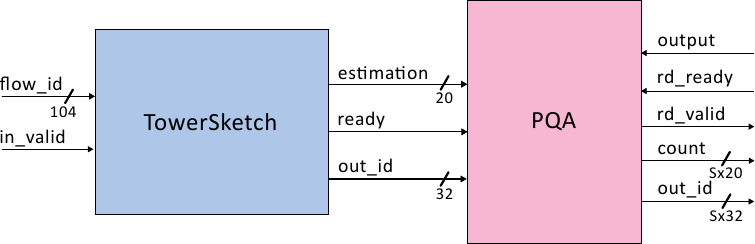}
		\caption{General architecture of the accelerator. The flow identifier of each packet is inserted into the TowerSketch, which estimates its frequency. The estimation is inserted into the PQA with a flow tag.}
                \label{general}
				\end{center}
    \end{figure}  
    
    Figure \ref{general} presents the general architecture of the accelerator, which implements Algorithm~\ref{alg:general}. The TowerSketch receives the 104-bit input \texttt{flow\_id}, which is constructed by concatenating five fields of the packet header: two 32-bit fields for the source and destination IP addresses, two 16-bit fields for the source and destination ports, and an 8-bit field for the protocol. The \texttt{in\_valid} signal indicates to the sketch that \texttt{flow\_id} contains a flow identifier for a valid packet that needs to be inserted. After inserting the packet, the sketch outputs an estimate of its flow frequency and asserts the output signal \texttt{ready}. When \texttt{ready} is high, the PQA updates its content with a tuple formed by frequency estimation and the 32-bit value \texttt{out\_id}, which is a 32-bit hash of \texttt{flow\_id}. Once all packets in the observation window have been processed, the PQA receives an \texttt{output} signal which starts the readout of its contents by the control plane. While \texttt{rd\_ready} is high, the PQA outputs the flow frequency and 32-bit hash \texttt{out\_id} for each of the $S$ flows in a row, and raises \texttt{rd\_valid}. This is repeated until the entire contents of the PQA are read.

    \subsection{TowerSketch architecture}

   \begin{figure}[tb]
	\begin{center}
	\includegraphics[width=1\linewidth]{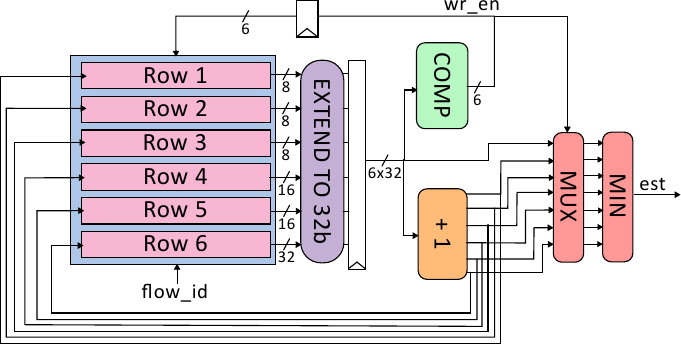}
		\caption{TowerSketch architecture. It reads the six counter values from memory blocks and extends them to 32 bits. It then compares them to determine which ones it must increment, and outputs a frequency estimation.}
                \label{towerarq}
				\end{center}
    \end{figure}  
    
    Figure \ref{towerarq} shows the architecture of the TowerSketch, which implements Algorithm~\ref{insalg}. As discussed in Section~\ref{sec:Tower}, the sketch consists of three rows of 8-bit counters, two rows of 16-bit counters, and one row of 32-bit counters. The \texttt{flow\_id} is used to read the value stored in a bucket of each row. To compare these values, we extend all of them to 32 bits, so if a bucket is overflowed, its extended value will be set to $2^{32}-1$. Each value is compared to all the others, generating the 6-bit \texttt{wr\_en} signal, where each bit is asserted only if the bucket of that row contains the minimum value and, therefore, must be updated. The \texttt{+1} block adds 1 to all buckets that are not overflowed. The incremented values are routed back to the sketch, but are only written if their corresponding  \texttt{wr\_en} signal is set. A multiplexer selects between the incremented and the old value for each bucket. Then, a comparator tree computes the minimum of these values to output the current flow frequency estimate.
    
    
    The memory section of the sketch, shown in blue in Fig.~\ref{towerarq}, is made up of 6 rows. In our implementation on an AMD UltraScale+ FPGA, each row is implemented using UltraRAMs of 4Kx64 bits.  The number of UltraRAMs used by each row depends on the total width of the rows $m=w_i\times \delta_i$, where $w_i$ is the number of counters in row $i$ and $\delta_i$ is the bit width of those counters. In our current implementation, $m=2^{21}$ bits, which requires 8 UltraRAMs per row. 
    
\begin{figure}[tb]
	\begin{center}
	\includegraphics[width=0.8\linewidth]{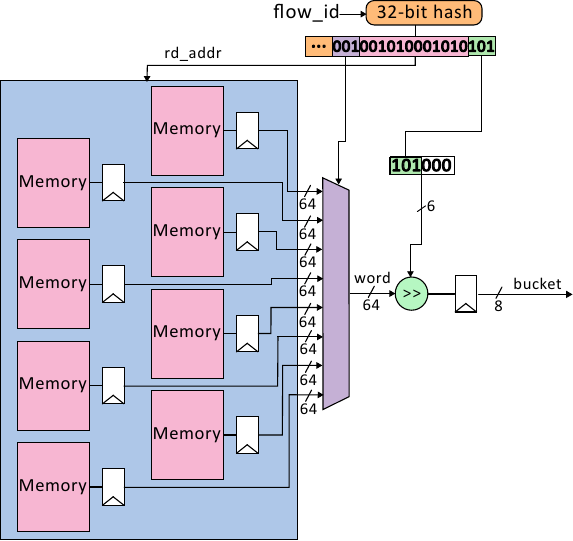}
		\caption{Read circuit for a row with 8-bit buckets. All memory blocks are read simultaneously and the 64-bit output is selected using a multiplexer. The counter value is extracted by shifting the word and keeping the least significant bits.}
                \label{reading}
				\end{center}
    \end{figure}

    As an example, Fig.~\ref{reading} shows the memory read circuit for an 8-bit bucket. When a packet is inserted, the sketch computes a 32-bit hash on the flow identifier. The lower 18 bits of the resulting hash are used to select an 8-bit counter from the sketch row: The 8 UltraRAM blocks are addressed using the middle 12-bits, thus selecting a 64-bit word from each of the blocks; the upper 3 bits select one of the 8 words; and the 3 least significant bits (LSBs) select one of the eight 8-bit counters that conform to the 64-bit word. The latter selection is performed by shifting the 64-bit result by a multiple of 8 bits between 0 and 56, and keeping the eight least significant bits of the result. For a 16-bit counter row, 17 bits of the hash are used, so that the 2 LSBs select one out of four 16-bit counters. For 32-bit counter rows, 16 bits of the hash are used, and the LSB selects one out of two 32-bit counters. In general, for a $2^{21}$-bit row of $\delta_i$-bit counters, the sketch uses $21 -\log_2 \delta_i$ hash bits to address the memory, of which 12 bits are used to address the eight UltraRAM blocks, 3 bits are used to select a 64-bit word, and $6-\log_2(\delta_i)$ bits are used to shift the word (in multiples of $\delta_i$ bits) to obtain the counter value. 
    
    After the counters are updated, the modified buckets are written back to the memory in a similar manner. The sketch inserts the modified value of the bucket into the previously read word, updating only the corresponding bits. Then, it writes this word into the selected memory, using the 12 middle bits of the hash as the memory address and the 3 upper bits as the UltraRAM-select signal.


    \subsection{PQA architecture}
            
    \begin{figure}[tb]
	\begin{center}
	\includegraphics[width=0.9\linewidth]{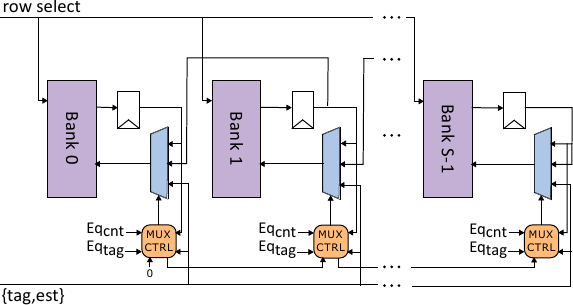}
		\caption{PQA update circuit. It uses a multiplexer per element to maintain the elements in a queue sorted.}
        \label{fig:pq_insertion}
	\end{center}
    \end{figure}  
The PQA module uses $S$ parallel memory blocks, each with $R$ entries. Upon every insertion, a subset of $\log_2 R$ bits of a hash function applied to the flow identifier is used to select an entry in each of the $S$ memory blocks (in practice, the accelerator reuses the output from the first hash function of the TowerSketch), forming a queue of $S$ elements. A combinational circuit then compares the rest of the hash of the input element and its frequency with all entries in the queue and updates the queue contents.

Figure~\ref{fig:pq_insertion} shows the PQA update circuit. Each memory block can receive one of three possible inputs: the input flow, an element coming from the memory block to the right (higher frequency), or the element already contained inside that memory block.
The logic considers three cases:
\begin{itemize}
    \item Case I: The input tag is not present in the queue and the input frequency is higher than the lowest one stored in it; therefore, it must be inserted in the queue. A signal $\texttt{Eq}_{cnt}$ encoded with a thermometer code is generated by comparing the frequencies of the elements in the queue to the incoming element, where a 1 indicates that the input element has a lower frequency. The 1-to-0 transition determines the insertion point. The input is inserted at this position and the lower-frequency elements shift accordingly, discarding the lowest-frequency element when the queue is full.

    \item Case II: The input tag is already present and must be updated with a higher frequency. The queue is kept sorted by shifting the elements that need to change position due to the updated frequency of the incoming element. This is determined by a second thermometer-encoded signal $\texttt{Eq}_{tag}$. An XOR operation with the $\texttt{Eq}_{cnt}$ signal determines which elements to shift.

    \item Case III: The input tag is not present, and its frequency is lower than that of all existing entries. No changes occur; the queue remains unmodified.
\end{itemize}

The PQA module uses two clock cycles to read the elements from memory, one to update the selected queue and another to write the queue back to the $S$ memory blocks. This process operates in a fully-pipelined fashion. To support continuous insertions, a two-cycle forwarding mechanism keeps an updated copy of memory content, allowing hazard-free operation when consecutive or near-consecutive flows access the same queue.

A finite-state machine controls the module, which transitions from insertion to reading upon input activation. When the contents are read out, the output interface delivers a full $S$-element queue per cycle with a valid signal, while simultaneously clearing the corresponding memory for reuse.

	\section{Experimental results}
    \label{sec:results}
    
\begin{table}[tb]
\centering
\caption{Network traces used in evaluation}
\label{traces}
\begin{tabular}{|l|r|r|}
\hline
\textbf{Trace} & \textbf{Flows} & \textbf{Packets} \\
\hline\hline
    Chicago-20150219-125911  & 635,775 & 15,808,577 \\
    Chicago-20160121-140200  &     866,722 & 31,166,491\\
    Chicago-20080319-200100  & 395,051 & 3,895,532 \\
   NYC-20181220-125909 &   2,339,880 & 29,633,594 \\
     NYC-20181220-130100  & 2,858,270 & 35,717,638 \\
     NYC-20190117-125910 & 2,426,848 & 29,492,979 \\
     NYC-20190117-130400  &      2,971,112 &37,921,931  \\
    NYC-20181220-140100  &         1,503,466 & 12,425,462 \\
     NYC-20190117-132100 & 7,338,987 & 83,104,571\\
    \hline
\end{tabular}
\end{table}

To evaluate our design, we used nine publicly available traces from the Center for Applied Internet Data Analysis (CAIDA) \cite{CAIDA}. These traces were collected from high-speed monitors in backbone links that connect Equinix to Chicago and New York. Table \ref{traces} presents the traces used in our evaluation, as well as the number of flows (cardinality) and the number of packets in each trace. All traces represent 1-minute measurements. Their cardinality ranges from 395 thousand to 7.3 million flows, and they contain between 4 and 83 million packets. We target top-$K$ detection with $K \leq 32,768$ (32K) to capture flows that account for at least 50\% of the traffic. 


\begin{table*}
\centering
\caption{Mean and standard deviation of ARE in the top-$K$ frequency estimates for different sketches}
\label{are}
\begin{tabular}{|c|r|r|r|r|r|r|r|r|r|r|}
\hline
\multirow{2}{*}{\textbf{$K$}} & \multicolumn{2}{c|}{\textbf{CMCU} \cite{goyal2012sketch}} & \multicolumn{2}{c|}{\textbf{CS} \cite{charikar2002finding}} &\multicolumn{2}{c|}{\textbf{Elastic} \cite{yang2018elastic}} & \multicolumn{2}{c|}{\textbf{TowerCU} \cite{yang2023sketchint}} & \multicolumn{2}{c|}{\textbf{Ours}} \\
\cline{2-11} 
& $\mu$ (\%) & $\sigma (\%)$ & $\mu$ (\%) & $\sigma (\%)$ &$\mu$ (\%) & $\sigma (\%)$
&$\mu$ (\%) & $\sigma (\%)$
&$\mu$ (\%) & $\sigma (\%)$\\
\hline
1K  & 0.11 & 0.07 & 0.25 & 0.11 & 1.27 &	1.23 & 0.05 &	0.08 & 0.01	& 0.01\\
2K  & 0.17 & 0.08 &0.31	&0.11 & 1.92 &	1.91 & 0.09 & 0.10 & 0.02& 0.02\\
4K  & 0.39 & 0.16 & 0.34&	0.12 & 3.01 &	2.36 & 0.22	& 0.21 & 0.05 &0.03\\
8K  & 1.01 & 0.42 &0.98 &	0.41 & 5.75 &	3.37 & 0.53	& 0.54 & 0.11 & 0.08\\
16K & 3.23 & 1.51 & 6.45 &	1.89 & 12.82	& 5.80 & 1.13 & 1.04 & 0.32	 & 0.20\\
32K & 15.86 & 7.13 & 35.19 & 11.60 & 39.25	 & 18.16 & 3.60 &	2.58 & 1.22&	0.73\\
\hline
\end{tabular}
\end{table*}

First, we compare the performance of our proposed TowerSketch with other count sketches used in hardware accelerators: Count-Min CU~\cite{saavedra2018heavy,soto2020hardware}, CountSketch~\cite{zazo2017single, fernandez2023streaming}, Elastic Sketch~\cite{yang2018elastic} and TowerCU~\cite{yang2023sketchint}.   Table \ref{are} compares the average relative error (ARE) in the frequency estimation of the top-$K$ flows for different values of $K$. To calculate ARE, we sort the top-$K$ frequencies estimated by each sketch, captured by a perfect priority queue of size $K$, and compare them with the real sorted frequencies of the top-$K$ flows. The sketches use approximately the same amount of memory: Both CountMin-CU and CountSketch use $12$ rows of $2^{15}$ 32-bit counters. TowerCU uses $3$ rows of 8-, 16-, and 32-bit counters, with each row using $2^{22}$ bits. Our modified TowerSketch uses $6$ rows, where each row uses $2^{21}$ bits. All these sketches use 1.54~MB of memory. Elastic Sketch uses a hash table of $2 \times K$ 85-bit entries and a CountMin-CU sketch with $13$ rows of $2^{14}$ 32-bit counters, for a total of 1.51~MB of memory usage. Each row in the table shows the mean and standard deviation of the ARE averaged over the 9 CAIDA traces. The table shows that, on average, our modified version significantly improves ARE with the same memory usage in all cases, which is even more notable for large values of $K$. For all sketches, the relative error increases with $K$. However, even for $K$ = 32K, our modified sketch achieves a mean ARE of 1.22\% on the 9 traces and less than 2.05\% on the trace with the worst performance.

\begin{table*}[tb]
\centering
\caption{Mean and standard deviation of top-$K$ precision for different sketches}
\label{preppq}
\begin{tabular}{|c|r|r|r|r|r|r|r|r|r|r|}
\hline
\multirow{2}{*}{\textbf{$K$}} & \multicolumn{2}{c|}{\textbf{CMCU} \cite{goyal2012sketch}} & \multicolumn{2}{c|}{\textbf{CS} \cite{charikar2002finding}} &\multicolumn{2}{c|}{\textbf{Elastic} \cite{yang2018elastic}} & \multicolumn{2}{c|}{\textbf{TowerCU} \cite{yang2023sketchint}} & \multicolumn{2}{c|}{\textbf{Ours}} \\
\cline{2-11} 
&  \multicolumn{1}{c|}{$\mu$}  & \multicolumn{1}{c|}{$\sigma $} &  \multicolumn{1}{c|}{$\mu$}  &  \multicolumn{1}{c|}{$\sigma $} & \multicolumn{1}{c|}{$\mu$}  &  \multicolumn{1}{c|}{$\sigma $}
& \multicolumn{1}{c|}{$\mu$}  &  \multicolumn{1}{c|}{$\sigma $}
& \multicolumn{1}{c|}{$\mu$}  &  \multicolumn{1}{c|}{$\sigma $}\\
\hline
1K  & 1.00 & 0.004 &  0.99 	& 0.009  &	 0.98 	& 0.018 &	 1.00 &	 0.003	& 1.00  &	0.000\\
2K  &    1.00 	&  0.003  &	 0.98 &	 0.019 &	0.97 &	 0.021 	& 1.00 	 &0.003 &	  1.00 &	0.000\\
4K  &  0.99 & 0.005 & 0.98 	& 0.015 &0.97 	& 0.022 &1.00 &0.006 &	 1.00 &	0.000
\\
8K  &  0.99 & 0.006 	&0.95 &	 0.019 &	0.95 	& 0.029 & 0.99 	& 0.012 	& 1.00 &	0.000
\\
16K &  0.97 & 0.013 & 0.85 & 0.035 & 0.91 &	 0.037 & 0.98  & 0.019  	& 1.00 &	 0.004
 \\
32K &  0.87 & 0.041 & 0.65 & 0.067 & 0.81 	& 0.047 & 0.96 	& 0.031 	 	& 0.99 &	 0.007
\\
\hline
\end{tabular}

\end{table*}

Table \ref{preppq} shows the precision achieved by the sketches in the identification of the top-$K$ flows. To calculate precision, we use as ground truth a 32-bit hash of the real top-$K$ flows, including all flows with the same frequency as the smallest top-$K$. We then compare this reference with the tags contained in the perfect priority queue (PPQ) of size $K$ that captures the output of the sketches. We calculate the precision as $P=\frac{TP}{TP+FP}$, where $TP$ (true positives) is the number of top-$K$ reported by the queue that were present in the ground truth, and $FP$ (false positives) are the flows that were not present in the ground truth. We observe that, among the sketches proposed in the literature, TowerCU obtains the best results with a minimum average precision of 0.96 for $K$ = 32K, which is significantly higher than the precision achieved by CountMin-CU (0.87), CountSketch (0.65) and Elastic Sketch (0.81). However, because our sketch redistributes the memory to increase the number of smaller counters, it matches or outperforms TowerCU for all values of $K$, with a minimum precision of 0.99 for $K$ = 16K, and 0.98 for $K$=32K.

\begin{table}[tb]
    \centering
    \caption{Mean and standard deviation of ARE in top-$K$ frequency estimates for different priority queues}
    \label{pq_are}
    \begin{tabular}{|c|r|r|r|r|r|r|}
        \hline
        \multirow{2}{*}{$K$} & \multicolumn{2}{c|}{PPQ} & \multicolumn{2}{c|}{PQA4 \cite{fernandez2023streaming}} & \multicolumn{2}{c|}{PQA6 (Ours)} \\ 
        \cline{2-7}
        & $\mu$ (\%) & 
        $\sigma$ (\%)  & $\mu$ (\%) & $\sigma$ (\%)  & $\mu$ (\%) & $\sigma$ (\%)  \\ \hline
        1K & 0.01  & 0.01  & 7.20& 1.26
  & 0.71  & 0.29  \\
        2K  & 0.02  & 0.02  & 8.32& 1.56
  & 0.78  & 0.17  \\
        4K  & 0.05  & 0.03  & 10.63	& 1.67
  & 1.14  & 0.26  \\
        8K  & 0.11  & 0.08  & 11.55	& 1.69
  & 1.28  & 0.26  \\
        16K & 0.32  & 0.20  & 12.75 &	1.85
  & 1.28  & 0.33  \\
        32K & 1.22  & 0.73  & 12.06	& 0.81
  & 0.88  & 0.23  \\
        \hline
    \end{tabular}

\end{table}

We now evaluate the performance of the PQA when capturing the top-$K$ flows from our TowerSketch frequency estimates. Table \ref{pq_are} compares the ARE in estimating the frequency of the top-$K$ flows for PPQ of size $K$, a PQA with $S=4$ and $R=K/4$ (PQA4), and a PQA with $S=6$ and $R=K/4$ (PQA6), which reflects our proposal of adding two extra buckets to each queue. In the latter case, after all the packets in the observation interval have been processed, we sort the PQA flows by frequency and keep the highest $K$ entries. Collisions in the hash function of the PQA4 cause the ARE to increase significantly with respect to the PPQ for all values of $K$, with values between 7\% and 13\%. In comparison, the PQA6 ARE are much smaller, varying between between 0.71\% and 1.28\%. In the trace with the worst performance, the ARE is less than 1.96\%.

\begin{table}[tb]
    \centering
    \caption{Mean and standard deviation of top-$K$ precision for different priority queues}
    \label{pq_p}
    \begin{tabular}{|c|r|r|r|r|r|r|}
        \hline
        \multirow{2}{*}{$K$} & \multicolumn{2}{c|}{PPQ} & \multicolumn{2}{c|}{PQA4 \cite{fernandez2023streaming}} & \multicolumn{2}{c|}{PQA6 (Ours)} \\ 
        \cline{2-7}
        &  \multicolumn{1}{c|}{$\mu$} & 
        \multicolumn{1}{c|}{$\sigma$}  & \multicolumn{1}{c|}{$\mu$} & \multicolumn{1}{c|}{$\sigma$}  & \multicolumn{1}{c|}{$\mu$}& \multicolumn{1}{c|}{$\sigma$}  \\ \hline
        1K &  1.00 & 	0.00&	 0.81 	& 0.01 	& 0.96 &	 0.01 
  \\
        2K  &  1.00 &	0.00 &	 0.81 &	 0.01 	& 0.95 	& 0.00 
\\
        4K  &  1.00 &	0.00	& 0.81 &	 0.01 	& 0.95 &	 0.00 
  \\
        8K  &  1.00 &	0.00	& 0.81 &	 0.00 &	 0.95 &	 0.00 
  \\
        16K &  1.00 &	 0.00 &	 0.81 &	 0.00 	 &0.95 &	 0.00 
 \\
        32K &  0.99 &	 0.01 &	 0.81 	& 0.00 	 &0.96 &	 0.01 
  \\
        \hline
    \end{tabular}

\end{table}

Finally, Table~\ref{pq_p} compares the average precision across all traces achieved by PPQ, PQA4 and PQA6 when identifying the top-$K$ flows from the TowerSketch estimates. The PPQ has almost perfect precision for all values of $K$. With PQA4, the precision drops to approximately 0.81 for all $K$ values, again as a product of hash collisions. Adding two additional elements to each queue in PQA6 significantly improves the mean precision to 0.95-0.96, with the worst trace achieving a precision of 0.94. These results show that PQA6 successfully approximates the behavior of PPQ, while PQA4 generally fails to adequately capture this behavior, mainly due to hash collisions when distributing flows across the queue array. The extra cost introduced by PQA6 compared to PQA4 is a 50\% memory overhead and the need to sort the queue contents after the observation interval. Despite this, PQA6 approximates the behavior of PPQ much more effectively than PQA4 and supports the insertion of one packet per cycle at high clock frequencies, which is impossible to achieve with the ideal PPQ.

\begin{table}[tb]
\caption{FPGA resource utilization}
\label{resourcefpga}
\begin{center}

\begin{tabular}{|c|r|r|r|r|r|}
\hline
 Resource & LUT & \text{FF} & URAM & \text{DSP}\\
\hline
\hline
TowerSketch & 7120 & 8054 & 48
 & 240\\
\hline
PQA & 1340 & 1414 & 12
 & 0\\
\hline
\hline
Total & 8460 & 9468 & 60 & 240\\
\hline
Utilization (\%) & 0.65 & 0.36 & 6.25 & 2.66 \\
\hline

\end{tabular}

\end{center}
\end{table}

We designed our accelerator at the register-transfer level using SystemVerilog and implemented it with Vivado 2021.1 on an AMD Alveo U280 board, featuring an XCU280 FPGA. We selected this board for its compatibility with NetFPGA PLUS, an open-source framework for smart network devices. Table~\ref{resourcefpga} reports resource utilization. The sketch uses 48 UltraRAMs for storage and the PQA6 uses only 12, totaling 6.25\% of the UltraRAM blocks of the device. The hash functions use 240 DSP blocks (2.66\%). Less than 0.7\% of the  LUTs are used to implement logic and distributed memory, while 0.36\% of the  flip-flops are used for pipeline registers and variables. The implementation runs at a clock frequency of 392~MHz on the selected device. The entire pipeline can process one packet per clock cycle, thus supporting a line rate of more than 200~Gbps even in the extreme case where all packets are minimum size (64 bytes). More realistically, the smallest average packet length in our test traces is 435 bytes, which would allow the accelerator to operate with line rates of 1300~Gbps using an input buffer. The proposed algorithm could also be implemented on P4-based programmable switches using previously proposed design techniques~\cite{soto2023sketch}.
	\section{Conclusions}
 
    We have presented an algorithm for the detection of top-$K$ flows in network traffic. We used a TowerSketch with conservative updates to estimate the flow frequencies, which we modified to accommodate the highly skewed distribution of the frequencies. We also use a data structure that approximates a priority queue, which keeps the most frequent flows and supports fast insertions in hardware. Our algorithm identifies top-$K$ flows in 9 real-world traces with a precision of more than 0.94 and estimates their frequencies with an average relative error of less than 1.96\%, outperforming previously proposed sketches and priority queue architectures. We designed a hardware accelerator for our algorithm and implemented it on an AMD Virtex XCU280 UltraScale+ FPGA. It runs at 392~MHz and operates at line rates of at least 200~Gbps. The accelerator uses less than 6.5\% of the device resources; therefore, it can be easily integrated with other network processing hardware on the data plane. In future work, we will leverage our results to estimate the statistical distribution of traffic based on the top-$K$ flows, and use it to estimate key data properties such as entropy and quantiles with low memory usage.

    \section*{Acknowledgements}

    This research was funded by the Chilean National Agency for Research and Development (ANID) through Fondecyt Grants 1220960 and 11240971, Basal Funds AFB240001 and AFB240002, and graduate scholarships 21191612 and 22231033.

    \bibliographystyle{IEEEtran}

    \bibliography{references}

\end{document}